\documentclass[slac_one]{revtex4}
\usepackage{graphicx}
\usepackage{fancyhdr}
\begin{document}

\title{Reconstruction and Identification of Hadronic Tau Decays with ATLAS} 

%

\author{S. Lai}
\affiliation{Albert-Ludwigs Universit\"{a}t Freiburg, 79104, Freiburg, Germany}
\author{A. Kaczmarska}
\affiliation{Institut of Nuclear Physics PAN, 31342, Krak\'ow, Poland}
\author{(on behalf of the ATLAS Collaboration)}

\begin{abstract}
The presence of $\tau$ leptons in the final state is an important signature in many 
Higgs boson and SUSY searches.  With the ATLAS detector, hadronically decaying
$\tau$ leptons can be reconstructed in a wide range of transverse energies.  
The reconstruction algorithm for hadronically decaying $\tau$ leptons is explained and the
performance of $\tau$ lepton identification is shown.  Particularly important is the 
discrimination of hadronically decaying $\tau$ leptons against overwhelming 
backgrounds from QCD jets.
\end{abstract}

\maketitle

\thispagestyle{fancy}


\section{INTRODUCTION} 

The $\tau$ lepton, with a mass of $1776.84\pm0.17$ MeV~\cite{PDG}, is the 
only lepton heavy enough to decay both leptonically and hadronically.  
It decays approximately 65\% of the time to one or more hadrons.
The reconstruction and identification of $\tau$ leptons are important in many 
searches for new phenomena, and they can appear in final states during the production 
of Higgs bosons or supersymmetric particles~\cite{CSC}.
This paper will discuss the reconstruction and identification of hadronically 
decaying $\tau$ leptons with the ATLAS detector~\cite{DetPap}.

\section{RECONSTRUCTION OF HADRONICALLY DECAYING $\tau$ LEPTONS}

Hadronically decaying $\tau$ candidates are reconstructed using at least one of two possible 
seed types.  The first seed type is a track with $p_T > 6$ GeV.  This leading track must 
satisfy further quality criteria to be considered a valid seed for a $\tau$ candidate.

The second type of seed consists of jets reconstructed using topological clusters (topoclusters)~\cite{CSC}
with $E_T > 10$ GeV.  Topoclusters are formed using cells that exceed calorimeter noise 
by 4$\sigma$, and neighbouring cells that exceed energy thresholds above calorimeter 
noise by 2$\sigma$ and 0$\sigma$ are associated to the cluster in a second and third step, respectively.  
These topoclusters are then grouped into a topojet using a seeded cone algorithm~\cite{conealg} with a 
cone radius of $\Delta R = \sqrt{(\Delta \eta)^2+(\Delta \phi)^2} = 0.4$ which form the 
aforementioned seeds for $\tau$ candidates.  
These topojets are then matched to seed tracks in a cone of radius $\Delta R = 0.2$ around the
topojet.   If such a match is found, the $\tau$ candidate is considered as having two valid seeds.
For reconstructed $\tau$ leptons in $Z\rightarrow\tau\tau$ events, 70\% of $\tau$ candidates have
two valid seeds, 25\% have only a topojet seed, and 5\% have only a track seed.

The energy of the $\tau$ candidate is calculated in two ways.  For $\tau$ candidates 
with a topojet seed, the cells in a cone of $\Delta R = 0.4$ are summed and weighted 
using a calibration similar to that used for the Liquid Argon calorimeter of the H1 
experiment~\cite{H1calib}.  Scale factors, determined from Monte Carlo and depending on  
the $p_T$ and $\eta$ of the $\tau$ candidate, are used to further correct the energy scale.

For $\tau$ candidates with a track seed, an energy flow approach is used, where energy 
deposits in cells matched to charged tracks are subtracted and 
replaced by the momenta of such tracks.  This energy of the $\tau$ 
candidate is also corrected for energy leakage coming from charged particles 
outside the narrow cone.

\begin{figure*}[h]
\centering
\includegraphics[width=70mm]{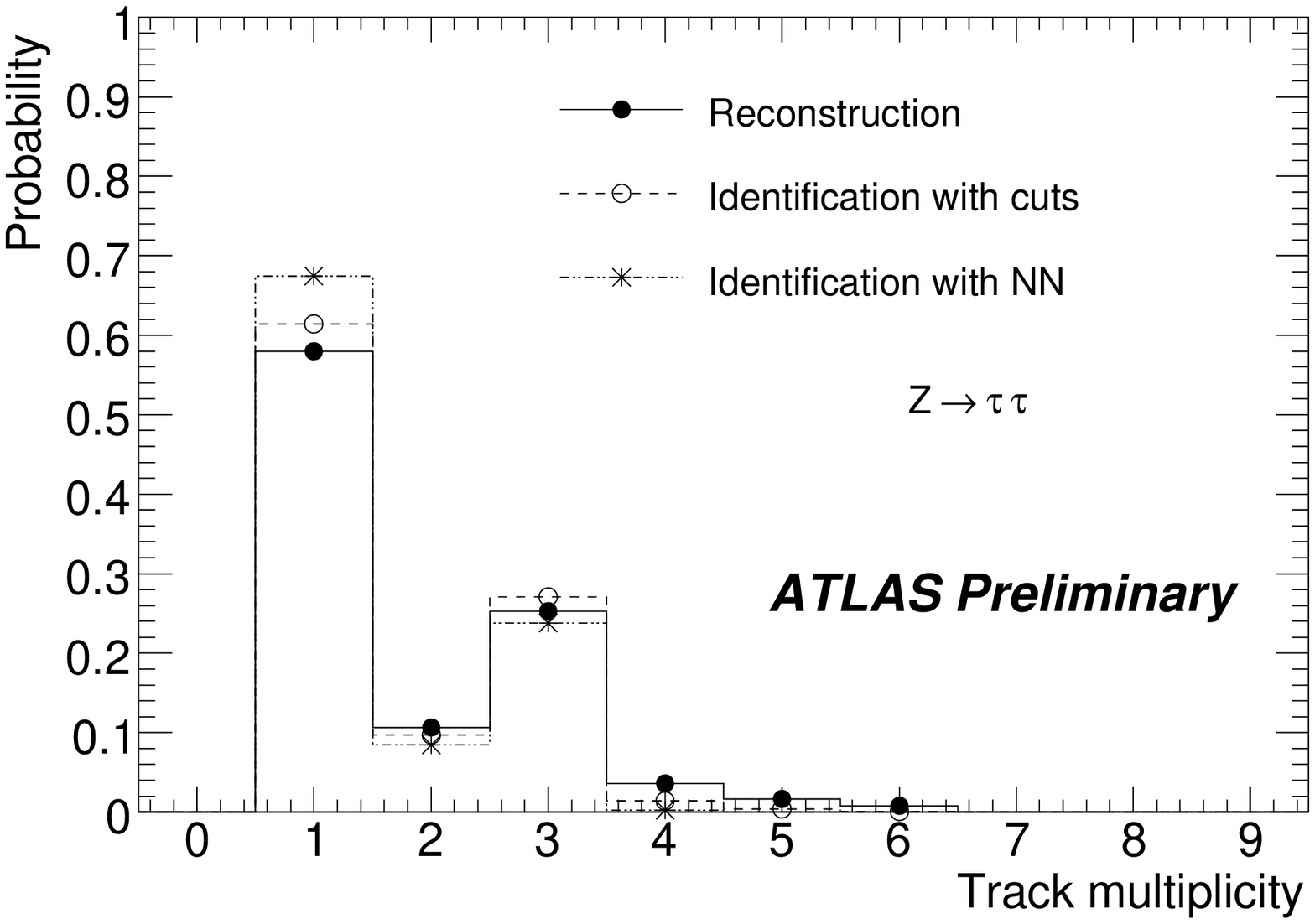}
\includegraphics[width=70mm]{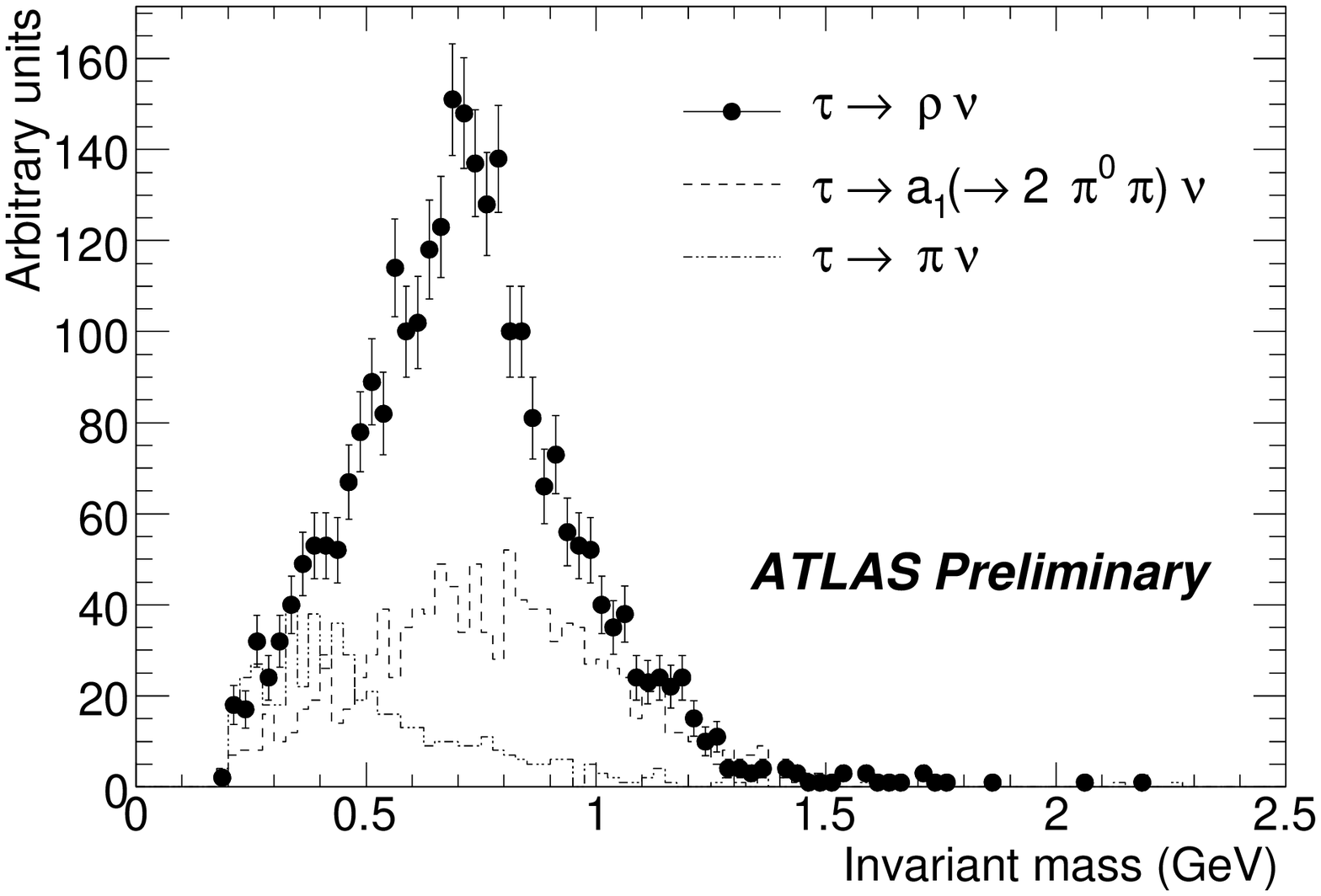}
\caption{Left: Track multiplicity for $\tau$ candidates in $Z\rightarrow\tau\tau$ events after reconstruction
             and identification steps.
             Right: Reconstructed invariant mass distribution for $\tau \rightarrow \pi \nu$, 
             $\tau \rightarrow \rho \nu$, $\tau \rightarrow a_1 \nu$ decays in $W\rightarrow \tau\nu$ events.} \label{track_dist_inv_mass}
\end{figure*}

Tracks associated to the $\tau$ candidate in a cone of $\Delta R = 0.2$ are required to pass 
track quality criteria, albeit less stringent than those for the leading track.  
Figure~\ref{track_dist_inv_mass} (left) shows the 
track multiplicity distribution for $\tau$ candidates matched to true hadronically decaying $\tau$ leptons.
Topoclusters found in the electromagnetic (EM) calorimeter with $E_T > 1$ GeV that are
isolated from tracks are interpreted as energy deposits from $\pi^0$ mesons in the 
$\tau$ lepton decay.  We find that approximately 66\% of $\tau \rightarrow \pi \nu$ 
decays are reconstructed with zero $\pi^0$ subclusters, while more than 50\% of 
$\tau \rightarrow \rho \nu$ ($\tau \rightarrow a_1 \nu$) decays are reconstructed 
with one (two) $\pi^0$ subcluster(s).

Based on the calorimeter information, the associated tracks and reconstructed $\pi^0$
clusters, a variety of other variables are calculated to be used for the identification 
of $\tau$ leptons.  These variables include the radius of the $\tau$ candidate
in the EM calorimeter, isolation 
variables for the calorimeter energy and tracks, reconstructed charge 
(based on the associated tracks), and the invariant mass of the $\tau$ candidate.  
The invariant mass distribution for selected $\tau$ candidates 
based on the above mentioned variables is shown in Figure~\ref{track_dist_inv_mass} 
(right).

\section{IDENTIFICATION METHODS FOR $\tau$ LEPTONS}

Because of the large production cross section for QCD jets, identification methods are needed
to discriminate $\tau$ candidates arising from true $\tau$ lepton decay and those from QCD processes.  
Identification methods are also used to distinguish hadronically 
decaying $\tau$ leptons from electrons and muons.  At ATLAS, various identification methods have
been studied, ranging from simple cut-based criteria, to multivariate techniques such as: projective
likelihoods (LH), boosted decision trees, neural networks, and probabilty range searches.
Here, only cut-based vetoes against muons and electrons and the projective
likelihood discriminant against QCD jets will be discussed.

Muons are vetoed by requiring that the calorimetric energy deposited by the $\tau$ candidate
has $E_T > 5$ GeV.  For electrons, cuts are placed upon 1-prong $\tau$ candidates on the following 
two quantities: the ratio of the transverse energy deposited in the EM calorimeter to the track transverse
momentum which tends to be higher for electrons than for charged hadrons, 
and the ratio of high threshold hits to low threshold hits in the Transition Radiation Tracker for the
track, which also tends to be higher for electrons.  
This veto suppresses electrons from $W\rightarrow e\nu$ events by a factor of 60, while retaining 
a 95\% efficiency for $\tau$ leptons in $W\rightarrow \tau^{had}\nu$.

\begin{figure*}[h]
\centering
\includegraphics[width=179mm]{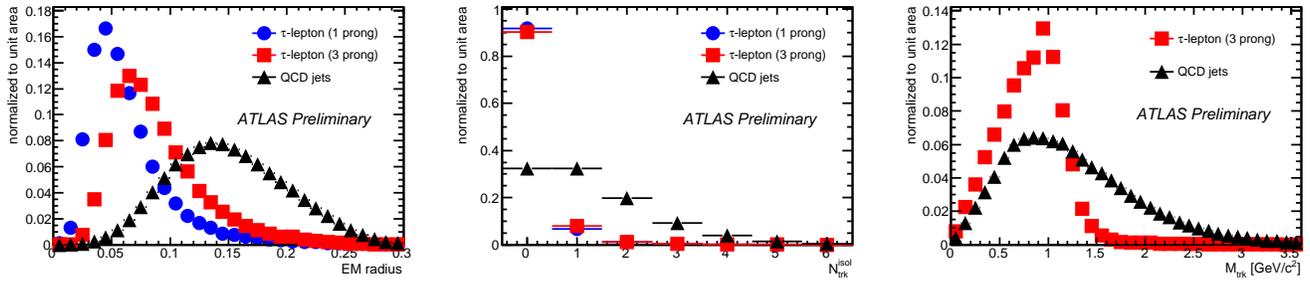}
\caption{Distributions of selected identification variables for 1-prong $\tau$ leptons,
3-prong $\tau$ leptons, and QCD jets.} \label{vardists}
\end{figure*}

Discriminating variables used to distinguish $\tau$ leptons from QCD jets include: 
radius and profile of EM calorimeter energy deposits, track width distributions, 
isolation variables calculated from calorimetric energy deposits and tracks, impact parameter significance
of the leading track, invariant mass of the associated tracks, ratios of energy deposits to the sum of track
transverse momenta, and the transverse flight path significance of the $\tau$ candidate vertex.
Sample distributions of a subset of these variables are shown in Figure~\ref{vardists}.

\begin{figure*}[h]
\centering
\includegraphics[width=73mm]{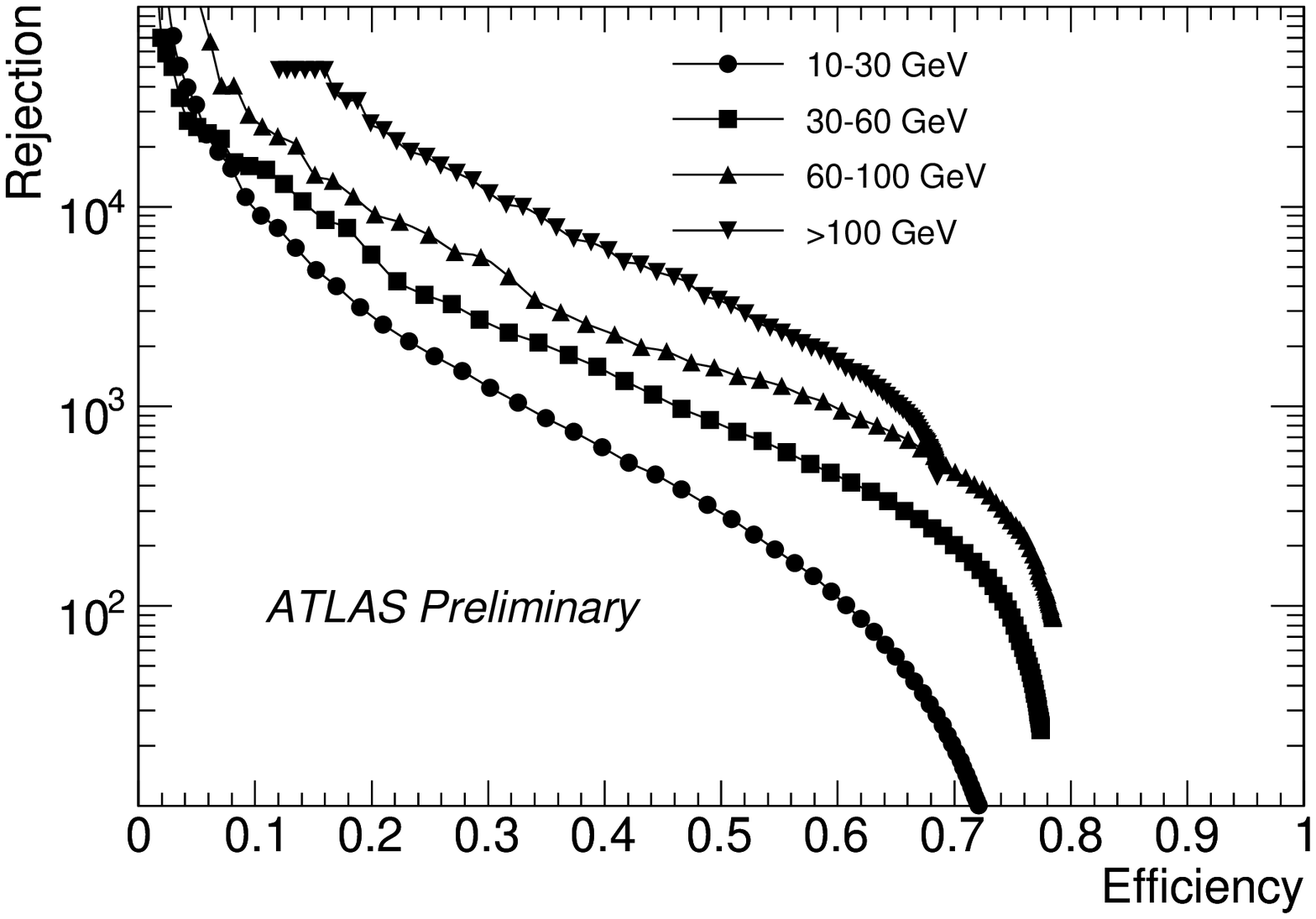}
\includegraphics[width=73mm]{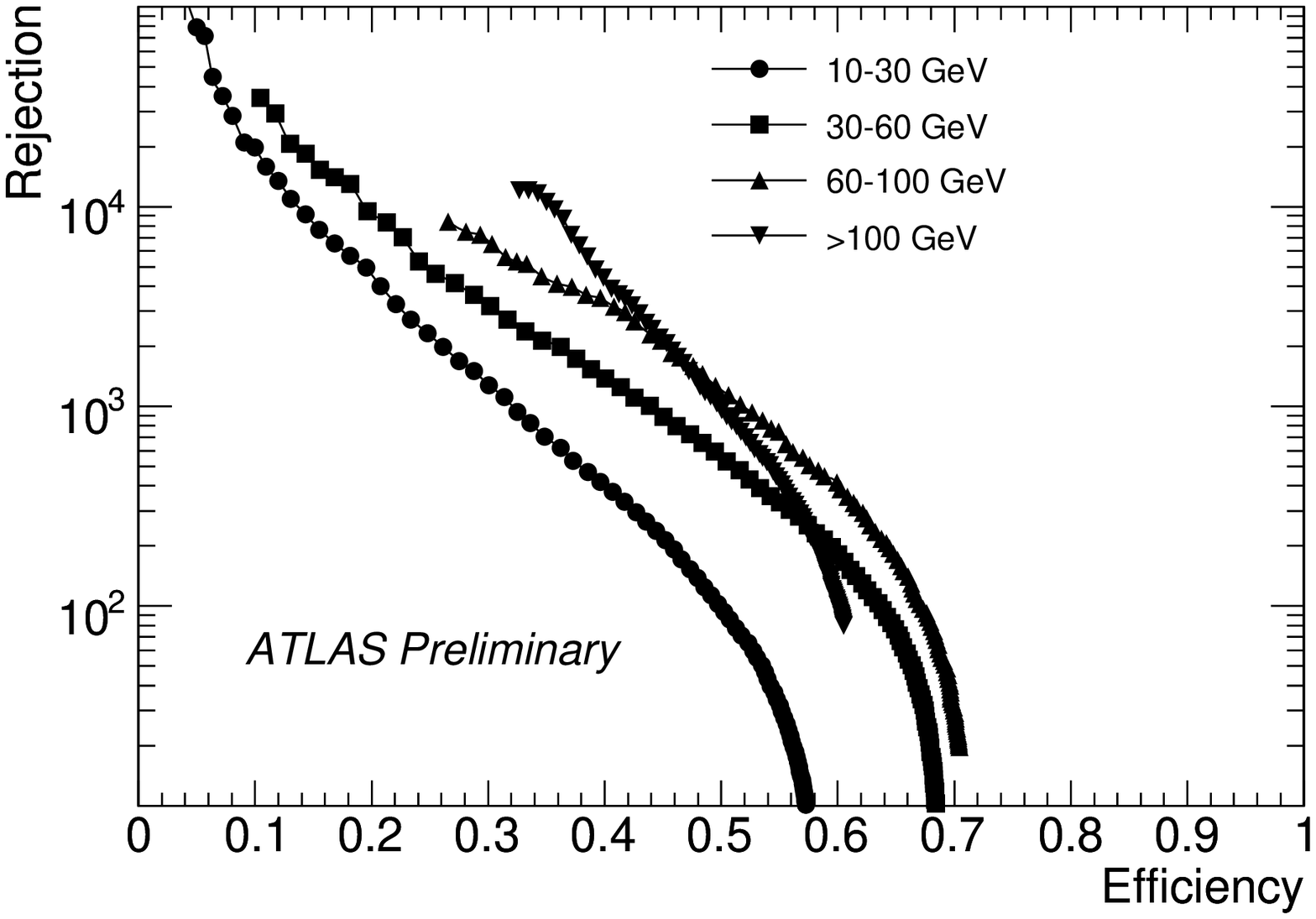}
\caption{Left: QCD jet rejection as a function of 1-prong $\tau$ efficiency for the projective likelihood discriminant.
             Right: QCD jet rejection as a function of 3-prong $\tau$ efficiency for the projective likleihood discriminant.} \label{LLHperf}
\end{figure*}

The LH takes these variables and forms probability distribution functions (PDFs)
for $\tau$ leptons (signal) and jets (background), which are separated into the following
categories: the $p_T$ of the $\tau$ candidate, the number of associated
tracks, valid seed types, and the presence of $\pi^0$ subclusters.  A likelihood ratio is formed using
the product of individual likelihood ratios for the signal and background hypothesis in each variable.  The performance
of the LH is shown in Figure~\ref{LLHperf} for 1-prong and 3-prong $\tau$ candidates.
For $\tau$ candidates with $30 < E_T < 60$ GeV, a 40\% $\tau$ identification efficiency corresponds to a
QCD jet rejection of $1500 \pm 100$ ($1480 \pm 100$) for 1-prong (3-prong) candidates.

\section{CONCLUSIONS AND OUTLOOK}

ATLAS is planning an extensive program to investigate physics channels with hadronically decaying $\tau$ leptons
in the final state.  Sensitivity to such channels requires robust reconstruction algorithms and well performing
identification methods to distinguish true $\tau$ leptons from QCD jets, electrons, and muons.  Studies for extracting
cross sections for Standard Model processes such as $W$, $Z$, and $t\bar{t}$ production with hadronic $\tau$ leptons 
in the final state have been 
performed~\cite{CSC}.  An understanding of these Standard Model processes will aid 
the understanding of searches for new physical phenomena with final state $\tau$ leptons.

\end{document}